\documentclass[pra,aps,twocolumn]{revtex4-2}
\usepackage{graphicx,amsmath,color,soul}

\newcommand{\PT}{\mathcal{PT}}

\usepackage[colorlinks,urlcolor=blue,citecolor=blue,linkcolor=blue]{hyperref}

\begin{document}

\title{Two-way enhancement of sensitivity by tailoring higher-order exceptional points}

\author{Ambaresh Sahoo}
\email{sahoo.ambaresh@gmail.com}
\affiliation{Department of Physics, Indian Institute of Technology Guwahati, Assam 781039, India}

\author{Amarendra K. Sarma}
\email{aksarma@iitg.ac.in}
\affiliation{Department of Physics, Indian Institute of Technology Guwahati, Assam 781039, India}

\begin{abstract}
Higher-order exceptional points in non-Hermitian systems have recently been used as a tool to engineer high-sensitivity devices, attracting tremendous attention from multidisciplinary fields. Here, we present a simple yet effective scheme to enhance the device sensitivity by slightly deviating the gain-neutral-loss linear configuration to a triangular one, resulting in an abrupt phase transition from third-order to second-order exceptional points. Our analysis demonstrates that the exceptional points can be tailored by a judicious tuning of the coupling parameters of the system, resulting in enhanced sensitivity to a small perturbation. The tunable coupling also leads to a sharp change in the sensitivity slope, enabling the perturbation to be measured precisely as a function of coupling. This two-way detection of the perturbation opens up a rich landscape toward ultra-sensitive measurements, which could be applicable to a wide range of non-Hermitian ternary platforms.
\end{abstract}

\maketitle

\section{Introduction}
\vspace{-0.3cm}
The non-Hermitian Hamiltonian with parity-time ($\PT$) symmetry was initially studied as a mathematical curiosity, and it resulted in a very different perspective of quantum mechanics \cite{Bender98, Bender02}. Today, there is hardly any branch of physics which is not influenced by these concepts \cite{El-Ganainy18, Christodoulides18}. The area which is most influenced both experimentally and theoretically is optics. In fact, optics turns out to be the test-bed for many key concepts related to parity-time symmetry. To date, experimental demonstrations of nonreciprocity \cite{Ruter10}, loss-induced transparency \cite{Guo09}, optical solitons \cite{Wimmer15}, $\PT$-symmetric microring lasers \cite{Hodaei14}, and so on \cite{Suchkov16} have been carried out using various $\PT$-symmetric optical platforms. Indeed, $\PT$-symmetry-induced effects have been experimentally observed in other branches of physics as well, such as electronics \cite{Schindler11}, mechanics \cite{Bender13}, atomic lattices \cite{Zhang16}, acoustics \cite{Shi16}, and Bose-Einstein condensates \cite{Kreibich14}. 

It is worthwhile to note that a non-Hermitian Hamiltonian $\hat{H}$ is said to be $\PT$ symmetric, provided $[\hat{H},\PT]=0 $. Here, $\mathcal{P}$ refers to the parity operator that simply interchanges two of the constituent modes of the system, while $\mathcal{T}$ is the time-reversal operator that takes $i\rightarrow -i$. One prominent aspect of such Hamiltonians is the breaking of $\PT$ symmetry, in which the eigenspectra switch from being entirely real to completely imaginary. Such an abrupt $\PT$ phase transition is marked by the presence of an exceptional point (EP) \cite{El-Ganainy18,Ruter10}. EPs of a $\PT$-symmetric Hamiltonian are the ones at which two or more eigenvalues and their associated eigenvectors coalesce simultaneously and become degenerate \cite{Miri19}. EPs open doors for completely new functionalities and performance. Nonreciprocal light propagation \cite{Peng14}, laser mode control \cite{Hodaei14, Feng14, Hodaei16, Peng16}, unidirectional invisibility \cite{Lin11, Longhi11, Regensburger12, Feng13}, optical sensing \cite{Wiersig14, Hodaei17, Chen17}, light stopping \cite{Goldzak18} and structuring \cite{Miao16}, and delaying the sudden death of entanglement \cite{Chakraborty19} are some notable phenomena triggered by EPs. Some additional features that are realized in optomechanics are phonon lasing \cite{Jing14}, optomechanically induced transparency \cite{Jing15, Lu18}, and sensitivity enhancement \cite{Liu16, Zhang20}. In condensed matter physics, connections between symmetries and EPs are investigated in two-dimensional (2D) systems \cite{Mandal21}.

Recently, higher-order EPs in an optical parity–time-symmetric coupled microresonator system have been experimentally demonstrated \cite{Hodaei17}, where a ternary $\PT$-symmetric photonic laser molecule with a carefully tailored gain-loss distribution is considered. The work reveals that the response of the ternary lasing molecule exhibits cube-root behavior, which could be used to improve the sensing performance of microresonator arrangements. In this paper, we consider different three-channel $\PT$-symmetric gain-neutral-loss configurations with linear and triangular arrangements of the multicore fibers. We work out the constraints under which such a system could be $\PT$ symmetric. We investigate how a small displacement of the neutral channel from the linear configuration changes the eigenvalues and the nature of the EPs. For that, we analyze the non-Hermitian system by introducing a triangular configuration. We find that the $\PT$-symmetric three-core structure exhibits higher-order EPs of order three that transform to order two with only a very slight deviation from the linear geometry. This abrupt phase transition affects the eigenvalue bifurcation diagram and the sensitivity to small perturbations, which could lead to ultrasensitive device applications. Furthermore, we investigate the device sensitivity to small perturbations in both linear and triangular configurations by tailoring the EPs, which leads to an efficient way of measuring small perturbations by simply changing the coupling.
%
\begin{figure}[t]
\begin{center}
\includegraphics[width=.475\textwidth]{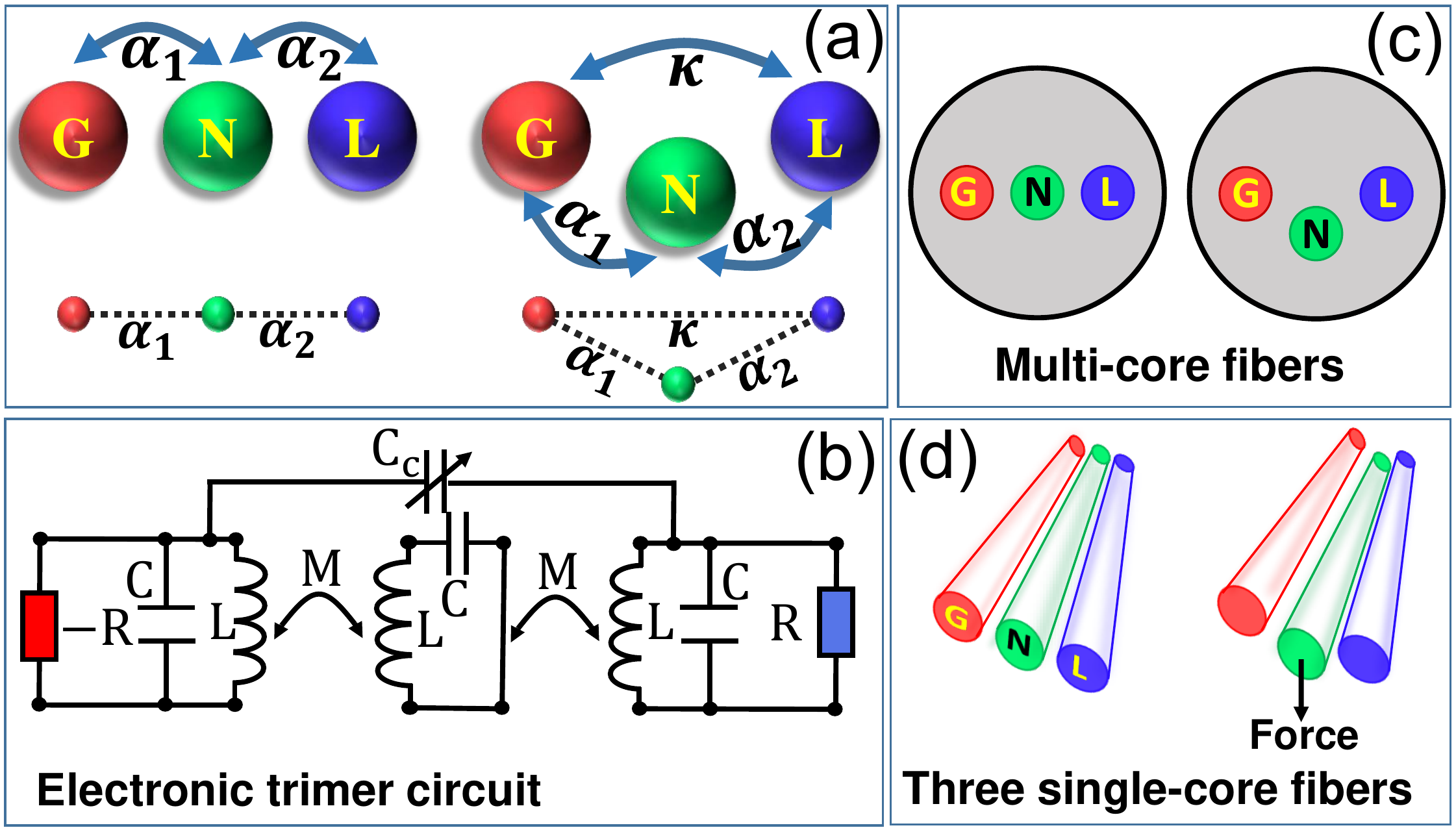}
\caption{(Color online) (a) Schematic diagram of three-channel couplers with linear and triangular configurations. Here, $\kappa$ is the coupling between the gain (G) and the loss (L) channels and $\alpha_1$ ($\alpha_2$) is the coupling between the neutral (N) channel and the gain (loss) channel. Different components of $\PT$-symmetric devices: (b) electronic trimer circuit (another possible configuration would be s third-order $\PT$-symmetric telemetry system \cite{Sakhdari19} with an additional coupling capacitor ${\rm C_c}$), (c) multicore fibers or PCFs, and (d) three closely spaced single-core fibers, where EPs can be tailored by controlling coupling (between gain and loss) for practical applications.}\label{fig:1model} 
\end{center} \vspace{-0.3cm}
\end{figure}

\vspace{-0.2cm}
\section{Description of systems}
\vspace{-0.3cm}
To describe the problem, we consider two possible types of three-channel $\PT$-symmetric structures with gain-neutral-loss configurations, as shown in Fig.\,\ref{fig:1model}. Here, the coupling coefficients can be tailored by changing the relative distance between the ports, and we get two possible configurations, namely, the triangular and linear configurations. Here, in general, there could be different coupling coefficients between two neighboring channels. However, in order for the system to be operated in the $\PT$ regime, these coupling coefficients must follow some relations, which we discuss in the next section. 

From a fabrication point of view, there are different fiber structures with three-core $\PT$-symmetric configurations that are easier to fabricate [either three cores surrounded by a common cladding or three solid cores in a photonic  crystal fiber (PCF) placed linearly/triangularly or a setup with three single-core fibers that are closely spaced and mutually coupled] than other configurations where loss/gain quadrupoles have been realized \cite{Christodoulides18, Hodaei17}. Traditionally, single-core fibers have been widely used over the years for various sensing applications that rely on mode switching, refractive index changes, and other fiber parameters that interact with the environment \cite{Joe18}. A $\PT$-symmetric electronic circuit \cite{Schindler11, Schindler12, Sakhdari19, Stegmaier21}  with three components can also be used to describe such ternary systems, where mutual inductance $M$ and a controllably coupled capacitor $C_c$ act as respective couplings. As an alternative, a variable coupled inductive circuit between the gain and loss components can be used as the coupling $\kappa$. 

For the generalized triangular configuration, shown in Fig.\,\ref{fig:1model}, one can construct the dimensionless coupled-mode equations that describe the evolution of field envelopes inside the first gain core ($a_g$), second neutral core ($b_n$), and third loss core ($c_l$) in the linear domain as 
\begin{align} \label{eqset1}
&i\frac{\partial}{\partial \xi}
\begin{pmatrix}a_g\\b_n\\c_l\end{pmatrix}
=\begin{pmatrix}
i\Gamma& \alpha_1 & \kappa\\
\alpha_1 & 0 & \alpha_2\\
\kappa & \alpha_2 & -i\Gamma
\end{pmatrix}
\begin{pmatrix}a_g\\b_n\\c_l\end{pmatrix}\equiv\hat{H} \begin{pmatrix}a_g\\b_n\\c_l\end{pmatrix},
\end{align}
where $\hat{H}$ represents the Hamiltonian of the system. $\xi$ is the dimensionless propagation distance, which could be time or any other parameter of the system under consideration. $\alpha_1$, $\alpha_2$, and $\kappa$ are the rescaled (by the characteristic length of the system) linear coupling coefficients between the neutral$-$gain ports, neutral$-$loss ports, and gain$-$loss ports, respectively. Also, $\Gamma$ is the rescaled balanced linear gain/loss coefficient. It is to be noted that when the coupling between loss and gain cores is zero, the triangular configuration becomes the well-known linear configuration \cite{Hodaei17}, and the corresponding pulse propagation equations are obtained by setting $\kappa=0$ in Eq.\,(\ref{eqset1}).

\vspace{-0.2cm}
\section{Exceptional points and their stability}
\vspace{-0.3cm}
The generalized Hamiltonian $\hat{H}$ describing both the linear and triangular coupled systems (schematically depicted in Fig.\,\ref{fig:1model}) will be $\PT$ symmetric if $\hat{H}$ satisfies the commutation relation $[\hat{H},\PT]=0$. It can be directly calculated from the commutation relation that, when $\alpha_1=\alpha_2=\alpha$, i.e., the couplings between the neutral port and the other two ports are equal, then the three-channel coupled configurations become $\PT$ symmetric. The complex eigenvalues corresponding to the $\PT$-symmetric Hamiltonian $H^{\PT}=\hat{H}$ (at $\alpha_1=\alpha_2=\alpha$) can be obtained by the direct method of diagonalization of the Hamiltonian as 
\begin{align}\label{eqset3}
\lambda_1=\frac{3^{1/3} f_1 + f_2^{2/3}}{3^{2/3} f_2^{1/3}},~
\lambda_{2,3}= -\frac{1}{2}\left(\lambda_1\pm i\frac{3^{1/3} f_1 - f_2^{2/3}}{3^{1/6} f_2^{1/3}}\right), 
\end{align}
where $f_1=2\alpha^2 +\kappa^2-\Gamma^2$ and $f_2= 9\alpha^2\kappa +\sqrt{81\alpha^4 \kappa^2 -3 f_1^3}$. Using this set of eigenvalues [Eq.\,\eqref{eqset3}], we analyze the bifurcation dynamics around EPs in detail.  We also investigate the stability and response of the system against a small perturbation around the EPs using the system Hamiltonian $\left(H^{\PT}\right)$. This analysis along with the transition from the linear to the triangular configurations give critical information with broad applications in non-Hermitian measurement devices with enhanced sensitivity.
\begin{figure}[t]
\begin{center}
\includegraphics[width=0.485\textwidth]{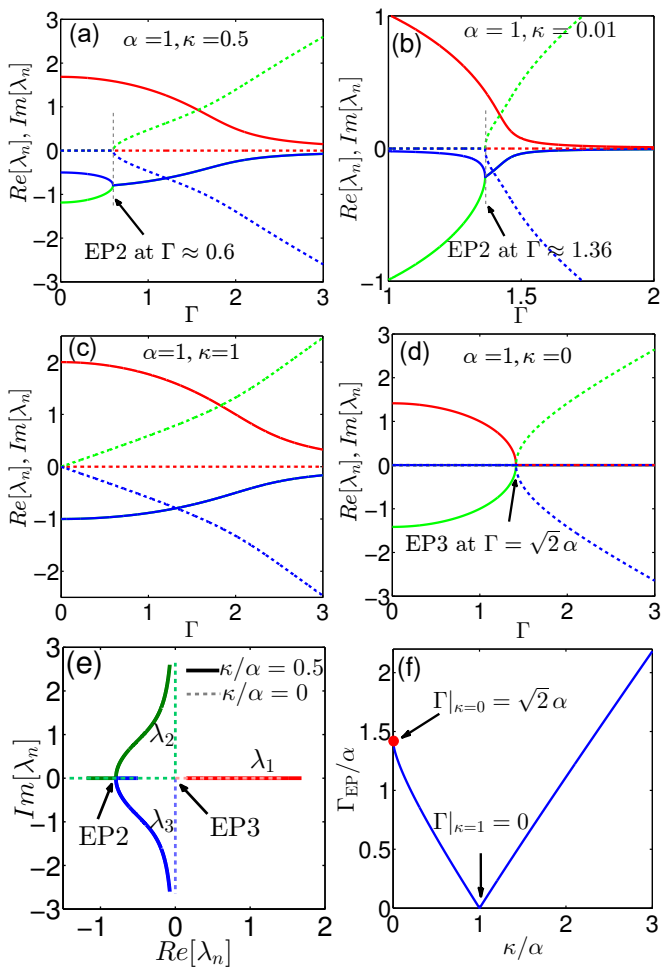}
\caption{(Color online) The real (solid curves) and imaginary parts  (dashed curves) of the eigenvalues ($\lambda_n$) as a function of the gain/loss parameter $\Gamma$ for $\mathcal{PT}$-symmetric (a)-(c) triangular and (d) linear configurations. The EPs [Eq.\,\eqref{eq4}] are denoted by arrows. (e) An alternative phase-space representation of the eigenvalues and EPs. (f) Manipulation of EPs by controlling the coupling contrast $\kappa/\alpha$. Here, the red circle represents the isolated fixed EP3.}\label{fig:2HEP} 
\end{center}  \vspace{-0.5cm}
\end{figure}

\vspace{-0.3cm}
\subsection{Tailoring exceptional points}
\vspace{-0.3cm}
To demonstrate the behavior of the three-channel $\PT$-symmetric system, we plot the eigenvalues as a function of $\Gamma$ in Figs.\,\ref{fig:2HEP}(a)-\ref{fig:2HEP}(d) for different sets of coupling coefficients. These eigenvalues could be energy or frequency spectra or any other physical parameters of the system. From Figs.\,\ref{fig:2HEP}(a), \ref{fig:2HEP}(b), and \ref{fig:2HEP}(d), it is evident that the eigenvalues are completely real below a specific threshold value of $\Gamma$, the EPs, denoted by arrows. Here, we observe that for triangular configurations ($\kappa \neq 0$), two of the three eigenvalues collapse to a single $\Gamma$ point (EP) from where the complex eigenvalues emerge. However, the third eigenvalue remains completely nonzero real. With increasing $\Gamma$, the real part of the degenerate eigenvalues and the nondegenerate real eigenvalues approach zero. In the case of a triangular configuration, this EP is a second-order EP (EP2), which, unlike the pure dipole $\PT$ case, does not coalesce to a point on the zero eigenvalue axis. It instead falls on a line perpendicular to the $\Gamma$ axis. This behavior is analogous to a two-atom system where a third atom acts as an impurity that changes the properties of the system. Now, in the case of a gain-neutral-loss linear quadrupole system ($\kappa=0$) shown in Fig.\,\ref{fig:2HEP}(d), this EP is a higher-order EP in nature, also known as a threefold EP (EP3). Unlike triangular configurations, here, only two real eigenvalues are nonzero that collapse at the EP3. From Eq.\,\eqref{eqset3}, the generalized analytical expression of the EPs on the $\Gamma$ axis can be calculated by equating $\lambda_2=\lambda_3$, implying $3^{1/3}f_1 -f_2^{2/3}=0$. The only physical solution for $\Gamma$ of this equation gives the EPs as
\begin{align} \label{eq4}
\Gamma_{\rm EP}=\alpha\,\sqrt{2 +\left(\frac{\kappa}{\alpha}\right)^2 -3 \left(\frac{\kappa}{\alpha}\right)^{2/3}}.
\end{align}
From Eq.\,\eqref{eq4}, one can get at $\kappa=\alpha$ there would not be any nonzero real eigenvalues, suggesting that the unbroken regime is forbidden when the system is an equilateral triangle in terms of coupling coefficients [refer to Fig.\,\ref{fig:2HEP}(c)]. It is to be noted from Eq.\,\eqref{eq4} that, in the absence of a neutral channel in between the gain and the loss channels, i.e., $\alpha=0$, the system becomes a standard two-core $\PT$-symmetric coupler for which the value of gain/loss strength at the EP2 is $\Gamma_{\rm EP2}=\kappa$ \cite{Miri19}. Thus, a specific choice of the coupling coefficients $\alpha$ and $\kappa$ directly controls the EPs of non-Hermitian systems. Alternative representations of the eigenvalues and EPs are shown in Fig.\,\ref{fig:2HEP}(e). Here, all three real versus imaginary parts of the $\lambda_n$ (thin light dashed lines) meet at the zero  crossing of the two axes (EP3) for the ternary linear arrangement ($\kappa=0$). However, for a triangular configuration with increasing $\kappa/\alpha$, the system abruptly changes its phase, and the three curves (broad  solid curves) move away from the center. In this case, the EP2 (indicated by an arrow) falls on the ${\rm Im}[\lambda_n]=0$ line, and can be tuned further by changing the coupling contrast $\kappa/\alpha$. Next, in Fig.\,\ref{fig:2HEP}(f) we plot the rescaled contrast of $\Gamma_{\rm EP}$ as a function of $\kappa$, implying that the system experiences a wide range of tunable EPs obtained by changing the geometry of the ternary alignment while maintaining isosceles triangular $\PT$ structures. 

It is to be noted that, for the linear configuration we have only one controlling parameter $\alpha$ apart from $\Gamma$, which results in an isolated fixed EP3, $\Gamma_{\rm EP3}/\alpha=\sqrt{2}$ [solid red circle in Fig.\,\ref{fig:2HEP}(f)]. However, with the triangular configuration, we have an additional controlling parameter $\kappa$ that tailors the EP2 [solid blue curve in Fig.\,\ref{fig:2HEP}(f)]. 
Also, in the limiting case of $\kappa/\alpha\gg 1$, the influence of the neutral core can be ignored, and the system behaves as a pure gain-loss binary $\PT$-symmetric system ($\Gamma=\Gamma_{\rm EP2}$). This limiting case can be confirmed by extending the plot of Eq.\,\eqref{eq4} [i.e., Fig.\,\ref{fig:2HEP}(f)] for larger $\kappa/\alpha$ values. The dynamics around EP2 and EP3 can also be efficiently tailored by deviating the system from its linear geometry, and the sensitivity to perturbations can be significantly enhanced through the sudden phase transition. Additionally, the strength of the perturbation can be precisely measured as a function of coupling coefficient contrast $\kappa/\alpha$, which we discuss below. It is to be noted that, near the EPs, the eigenvalues of the system as a whole behave as a perturbed EP3 system for $\kappa/\alpha\ll 1$. This critical behavior is entirely different from that of the pure EP2 system, which contributes to the sensitivity enhancement.

\vspace{-0.3cm}
\subsection{Stability against small perturbations}
\begin{figure}[t]
\begin{center}
\includegraphics[width=0.445\textwidth]{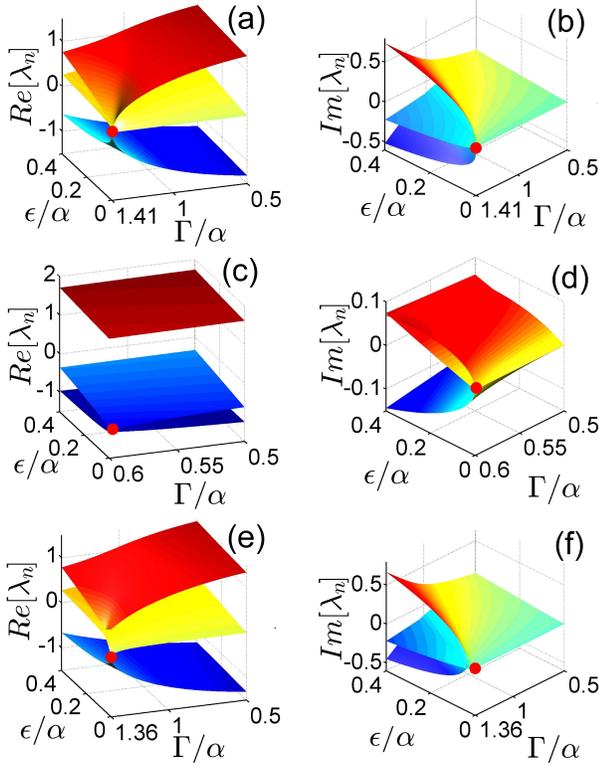}
\caption{(Color online) Bifurcation of complex eigenvalues around EPs in the presence of perturbation: (a), (b) for a linear configuration ($\kappa/\alpha=0$); (c), (d) and (e), (f) for triangular configurations with $\kappa/\alpha=0.5$ and $\kappa/\alpha=0.01$, respectively. Here, the solid red dots represent the EPs.}\label{fig:3} 
\end{center} \vspace{-0.5cm}
\end{figure}

\vspace{-0.3cm}
To demonstrate the sensitivity of the system against perturbations, we choose the system to be operated near the EP3 by setting the controlling parameters. We introduce a small perturbation $\epsilon$ imposed on the gain channel. In doing so, the Hamiltonian of the system is modified as
\begin{align}  \label{eq5}
H'=\begin{pmatrix}
i\Gamma+\epsilon& \alpha & \kappa\\
\alpha & 0 & \alpha\\
\kappa & \alpha & -i\Gamma
\end{pmatrix}. 
\end{align}
This perturbation could, in principle, be introduced in any other port. In order to understand how the perturbation changes the dynamics of the system near the EPs, we first consider the linear configuration. In Fig.\,\ref{fig:3}(a) and \ref{fig:3}(b) the bifurcation diagram of the complex eigenvalues around an EP3 (indicated by solid red circle) is plotted from Hamiltonian $H'$ [Eq.\,\eqref{eq5}] with a rescaled gain/loss contrast and detuning $\epsilon/\alpha$ for $\kappa=0$. 
%

\begin{figure}[t]
\centering
\begin{center}
\includegraphics[width=.485\textwidth]{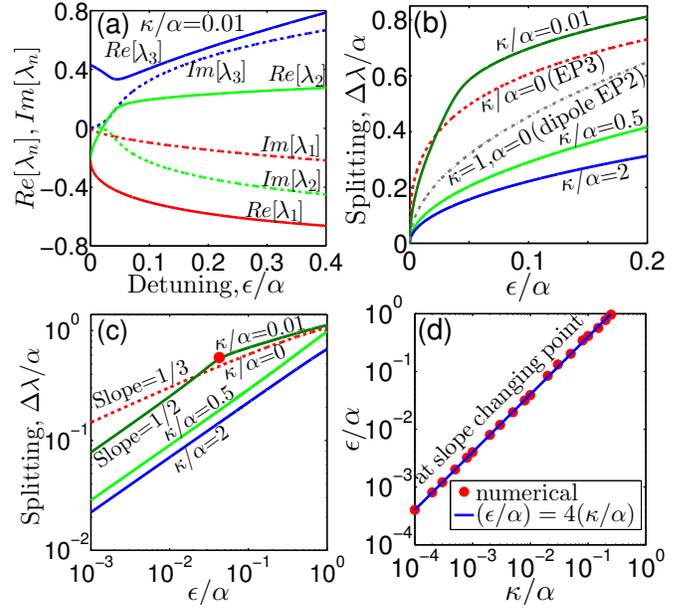}
\caption{(Color online) (a) Evolution of real (solid curves) and imaginary (dashed curves) parts of the eigenvalues around the EP as a function of perturbation. Sensitivity in terms of the splitting of eigenvalues ${\rm Re}(\lambda_2-\lambda_1)/\alpha=\Delta\lambda/\alpha$ against perturbations for different coupling contrasts ($\kappa/\alpha$) in (b) linear and (c) logarithmic scales. (d) Precise measurement of the strength of perturbation by locating the slope changing point ($1/2\rightarrow 1/3$) [solid red circle in (c)] as a function of $\kappa/\alpha$.}\label{fig:4} 
\end{center} \vspace{-0.5cm}
\end{figure}

Next, we consider the generalized case with $\kappa\neq 0$. We plot the bifurcation diagram of the eigenvalues around the EP2 (solid red circles) in the presence of perturbations in Fig.\,\ref{fig:3}(c)-\ref{fig:3}(f). As we mentioned earlier, for the triangular case, the EP2 falls on a line perpendicular to the $\Gamma/\alpha$ axis. The real parts of the two eigenvalues fall on the EP2 line, while their imaginary parts coincide at $\lambda_n=0$ on the line. Figures\,\ref{fig:3}(c) and \ref{fig:3}(d) also show that, unlike the linear configuration, here, the imaginary eigenvalue surfaces are more wrapped in the presence of perturbations, and even at some point, they intersect each other. From these bifurcation diagrams [Figs.\,\ref{fig:3}(e) and \ref{fig:3}(f)], the variation of the eigenvalues at the EP2 with respect to the perturbation are plotted in Fig.\,\ref{fig:4}(a), where $Re[\lambda_n]$ values (solid curves) show a trend more as the linear configuration (the case with EP3) as one moves away from the EP2. However, when the system is observed close to the EP2, it behaves differently, and one of the real eigenvalues suddenly changes its curvature while remaining nondegenerate. 

To check how the system reacts to the perturbation around the EP2, we plot the sensitivity in Figs.\,\ref{fig:4}(b) and \ref{fig:4}(c), i.e., the difference between two eigenvalues ${\rm Re}(\lambda_2-\lambda_1)$ in linear and logarithmic scales. The splitting (sensitivity) can be improved by tuning the $\kappa/\alpha$, as shown in Fig.\,\ref{fig:4}(b). In this case, we observe an enhancement in splitting at $\kappa/\alpha<1$ toward smaller detuning and at $\kappa/\alpha>1$ toward relatively larger detuning (which can be verified with $\kappa/\alpha\ge 5$, not shown here), with $\kappa/\alpha=1$ providing the lowest bound because this particular coupling contrast results in the nonexistence of an EP [shown in Figs.\,\ref{fig:2HEP}(c) and \ref{fig:2HEP}(f)]. It is worth noting that these results are physically consistent with the fact that the splitting of eigenvalues in the vicinity of EP3 is greater than that of EP2 for $\epsilon/\alpha\ll 1$. However, when an additional coupling or interaction is introduced, we see more splitting in some regions (beyond EP3 dominance), which is unique for mixed-EP systems. The overall findings are remarkable in this case. First, by changing the coupling contrast from zero to nonzero, we can change the bifurcation dynamics of the system that leads to the transition from EP3 to EP2. In doing so, the sudden change in the dynamics of the system can be detected through its eigenvalues, which is further controlled in the case of EP2 by changing the coupling contrast. Second, the transition from EP3 to EP2 also changes the sensitivity from $\sim\epsilon^{1/3}$ to $\sim\epsilon^{1/2}$ and vice versa. Furthermore, for a small coupling contrast, i.e., $\kappa/\alpha< 1$, there is a sharp change in slope in the logarithmic plot [Fig.\,\ref{fig:4}(c)] from $1/2$ to $1/3$,  with a case $\kappa/\alpha=0.01$ represented by a solid red circle. This slope changing point is further plotted as a function of coupling contrast in Fig.\,\ref{fig:4}(d) with solid red circles, which follows a linear relation from the fitting: $\epsilon/\alpha = 4\,\kappa/\alpha$ (solid blue curve). These findings are important because a slight variation of the coupling contrast $\kappa/\alpha$ on the order of $10^{-4}$ or less could make the transition from EP3 to EP2 of the non-Hermitian system and also change the sensitivity slope from $1/3$ to $1/2$ even for very small perturbations of the same order of magnitude as $\kappa/\alpha$. This way, a very small perturbation can be precisely measured by tuning the coupling contrast and looking at where the transition from one slope to the other takes place. This controllable $\PT$-symmetric system with EP manipulation could be applied in high-sensitivity measurement devices to detect a very tiny perturbation $\epsilon$ by slightly deviating the linear configuration through the controlling parameter $\kappa/\alpha$. This ternary system could also be useful in measuring forces through deflection of the neutral element, as EP and the deflection of the neutral component are directly related.

\vspace{-0.3cm}
\section{Discussions}
\vspace{-0.3cm}
In this paper we have studied how one can obtain excellent sensitivity against perturbations in a $\PT$-symmetric three-channel system through manipulation of higher-order EPs. We demonstrate that a tiny deviation from the linear to triangular configuration causes the phase transition of the system, resulting in a transition from EP3 to EP2. In addition, the sensitivity of the system to a small perturbation can be enhanced by tuning the coupling contrast. Furthermore, the coupling contrast causes a sharp change of the slope of the sensitivity curve from $1/2$ to $1/3$. This two-way change of the system dynamics can be utilized to enhance the device sensitivity and efficiently measure the strength of perturbation. These findings could pave the way for manipulating the system dynamics and improving sensitivity in various non-Hermitian device applications, potentially opening up different opportunities in other interdisciplinary fields. 

We may think of a few possible experimental scenarios in which our proposed idea of tailoring EPs may be applicable. In $\PT$-symmetric electrical circuits \cite{Sakhdari19,Stegmaier21} and circuit-QED systems \cite{Baust15,Quijandria18}, the three-component coupled circuit (shown in Fig.\,\ref{fig:1model}) can be utilized, whose coupling contrast can be tuned by controlling the coupled capacitor $C_c$. Other systems include optomechanics \cite{Jing17,Xiong21}, atomic systems \cite{Zhang16,Sheng13}, acoustic vibration of membranes attached to mutually coupled inductances of a $\PT$-symmetric electrical circuit, and so on. A system consisting of three closely aligned single-core mutually coupled fibers (schematically depicted in Fig.\,\ref{fig:1model}), waveguides, and resonators with the gain-neutral-loss configuration can be utilized for deflection or stress measurement. Additionally, gain-neutral-loss cores with linear and triangular alignments with different sets of $\kappa/\alpha$ can be fabricated in a single PCF or as multi-core fibers \cite{Longhi16} for various sensing applications.


\end{document}